# A NEW SELF-PROPELLED MAGNETIC BEARING WITH HELICAL WINDINGS

**B Shayak**

Department of Theoretical and Applied Mechanics,
School of Mechanical and Aerospace Engineering,
Cornell University,
Ithaca – 14853,
New York, USA

sb2344@cornell.edu , shayak.2015@iitkalumni.org

✳

## Abstract

In this work a design is proposed for an active, permanent magnet based, self-propelled magnetic bearing i.e. levitating motor having the following features : (*a*) simple winding structure, (*b*) high load supporting capacity, (*c*) no eccentricity sensors, (*d*) stable confinement in all translational dimensions, (*e*) stable confinement in all rotational dimensions and (*f*) high efficiency. This design uses an architecture consisting of a helically wound three-phase stator, and a rotor with the magnets also arranged in a helical manner. Active control is used to excite the rotor at a torque angle lying in the second quadrant. This torque angle is independent of the rotor's position inside the stator cavity hence the control algorithm is similar to that of a conventional permanent magnet synchronous motor. It is motivated through a physical argument that the bearing rotor develops a lift force proportional to the output torque and that it remains stably confined in space. These assertions are then proved rigorously through a calculation of the magnetic fields, forces and torques. The stiffness matrix of the system is presented and a discussion of stable and unstable operating regions is given.

✳    ✳    ✳    ✳    ✳

# Nomenclature

Vectors are denoted upright i.e. non-italicized and bold e.g. **A**. All vectors are in real three-dimensional space – no electrical phasors have been used. The magnitude of **A** is denoted by $A$. Unit basis vectors are denoted by a hat above the corresponding coordinate for example $\hat{\mathbf{x}}$. Components of a vector are denoted by single subscripts e.g. $A_x$. Matrices are denoted upright, bold and enlarged e.g. **C**. Overhead dots are used to indicate the derivative with respect to time, thus $\dot{x}$.

The following is a list of the most important variables used in this Article :

- **B** : the magnetic field vector
- **F** : the force vector
- $h$ : half the height of the motor
- $I$ : the transverse moment of inertia of the rotor
- $i$ : the wire current in the secondary windings
- **K** : the surface current (current per unit length) in the primary windings/magnets
- $k$ : the axial wavenumber of the windings
- $m$ : the mass of the rotor
- $n$ : the semi-polarity of the windings
- $R$ : the stator radius
- $r$ : the rotor radius
- $(\ldots)_r$ : related to the rotor. This subscript appears before all other subscripts
- $(\ldots)_s$ : related to the stator. This subscript appears before all other subscripts
- **T** : the torque vector
- $x_{CM}, y_{CM}, z_{CM}$ : the displacement of the rotor centre of mass relative to the geometrical centre of the stator, measured in stator coordinates
- $\alpha$ : the pitch angle of the helical windings. $\alpha = \arctan kR$
- **κ** : the system stiffness matrix
- $\theta$ : the angle of nutation of the rotor
- $\delta$ : the torque angle
- $\eta$ : the gradient in $\mathbf{B}_s$

This completes the definitions of notations used systematically throughout the Article. Due to the length and complexity of the calculations, many variables appear transiently – for example an Eulerian coordinate basis $a,b,c$. These sporadic usages have not been listed in the above.

<div align="center">✽</div>

# 1 Introduction and the Proposed Design

This Section contains three Subsections. The zeroth Subsection summarizes the literature and sets the context for the present work. The first Subsection describes the design of the proposed magnetic bearing, in completely qualitative terms. The second Subsection motivates on purely physical grounds the hypothesis that the proposed design might indeed support the weight of a heavy rotor and show stable magnetic confinement.

## 1.0 Introduction

Magnetic bearings are, at least in theory, an excellent means of supporting heavy and high-speed rotating shafts. Since the bearings are non-contact, there are no stresses induced and hence no possibility of wear and tear and/or fatigue failure. In appliances such as washing machines where the rotors are eccentric by definition, conventional bearings nearly always fail after a certain duration of usage due to the cyclic stresses; such failures are often the driving factor behind the overall life of the appliance. These and similar issues can be rendered moot if magnetic bearings are employed instead of conventional bearings. Nevertheless, a widespread application of magnetic bearings is still not present today, primarily due to the challenges involved in designing a bearing which can combine the various qualities desired from it.

There are two types of magnetic bearings prevalent in the market today : active and passive. Active bearings generally use permanent magnets on the rotor, thus allowing for the creation of strong magnetic fields which can generate large forces. The drawback of these systems is that they require continuous monitoring of the rotor eccentricity, which is processed through a suitable controller to regulate the current in the various windings of the stator. Achieving the required precision makes the system prohibitively expensive, and also prone to catastrophic failure in the event of a fault in the eccentricity sensors and/or controller. Some treatments of this may be found in [1-4].

Passive magnetic bearings achieve the confinement of the rotor without using eccentricity dependent control. This implies a much lower cost and much higher reliability over the active magnetic bearing case. Some designs use permanent magnets while others use eddy current mechanisms and are called electrodynamic bearings. In permanent magnet bearings, there is often, though not always, one separate set of magnets installed for each degree of freedom which is to be confined [5,6]. This can lead to a multiplicity of coils and magnets being required for a relatively simple application. Further, due to Earnshaw's theorem, a shaft cannot be supported through passive permanent magnet bearings alone [7]. A novel configuration called whirl imposer invented by QINGWEN CUI, JAN SANDTNER and HANNES BLEULER [8] bypasses this restriction by converting the instability of the permanent magnets into precession of the rotor, as in the toy called Levitron [9,10]. For some applications however, the finite nutational angle in this mode of operation might be a drawback. In electrodynamic bearings, single windings often suffice to achieve confinement of multiple degrees of freedom, as in homopolar bearings [11-13]. Further, since eddy currents are exempt from Earnshaw's theorem by default, no active control is theoretically required to support a shaft on these bearings. Unfortunately, eddy currents typically produce much weaker fields than permanent magnets and hence can support smaller loads. A second disadvantage of these bearings is that the resistive losses in many designs is considerably high compared to the input power of the device being supported.

A typical magnetic bearing application has been considered by GERALD JUNGMAYR et. al. [14] in an article where they describe a compact (miniature) fan mounted on magnetic bearings. All the bearings involve permanent magnets – there are two passive and one active bearing, each with its own set of magnets and coils. Over and above this, there is the powerplant i.e. the motor itself. This is quite a complex arrangement for a small and light apparatus which the authors want to operate. This complexity is one factor which restricts these bearings to speciality applications. One step towards wider applicability can be achieved if it is possible to reduce the number of separate windings and magnets required to achieve confinement of the rotor. Further simplification can be achieved by using passive systems for control of as many degrees of freedom as possible.

In this work, a design of a self-propelled permanent magnet synchronous motor (PMSM) bearing is proposed which can support a high gross weight. The bearing uses one principal and one auxiliary set of windings and magnets – the principal set, which provides propulsion, weight support and a substantial fraction of the confinement, is operated exactly as in a conventional (non-levitating) PMSM and the auxiliary windings are simply excited with dc. The control of the torque angle is the only active step required for levitation and stabilization of the rotor. The analytic methods used to prove these conclusions have their origins in the works of VIRGINIE KLUYSKENS et. al. and JOAQUIM DETONI et. al. [15-22]. The approach found in some of these works – that of starting from electromagnetic field descriptions as opposed to a lumped parameter equivalent circuit model – is extended in the present Article to obtain the stable regions in the parameter space of the proposed system.

## 1.1 The Design

In this Subsection the design of the proposed bearing is described. It consists of a cylindrical stator and a cylindrical rotor suspended freely inside it. The normal operating state features the rotor concentric and coaxial with the stator. Gravity acts along the axis of the stator. There are two sets of windings on the stator matched by corresponding magnets and/or windings on the rotor – one principal and one auxiliary. The principal stator winding is three-phase, like in a conventional PMSM. The phase windings however are not parallel to the axis of the cylinder but are twisted into the shape of a helix. A schematic diagram of this is shown in Fig. 1. The pitch of the helix depends on the application requirements. The polarity is arbitrary, again dependent on the application. The stator may be supplied with either a voltage source or a current source inverter – continuous variation of the stator current is mandatory.

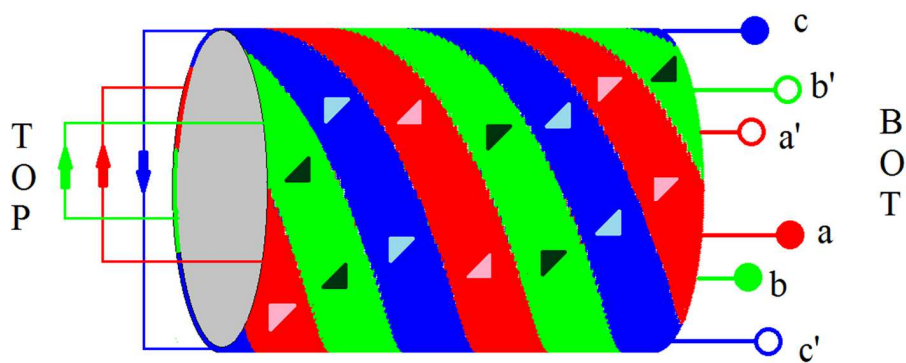

Figure 1 : *The stator. This figure shows the polarity as 2 – the most basic configuration. In reality the stator is mounted vertically, shown by the labels 'top' and 'bot'. a,b,c denote the three phases – for a positive phase current, the proper branch carries the upward current while the primed branch carries the downward current. These directions have been shown with arrows. It is to be noted that the winding arrangement is identical to that of a conventional motor, the only extra feature being the helicity. The windings are right handed; over the height of the stator, all branches perform one full turn.*

On the rotor, permanent magnet pole pieces are arranged on the surface, also in a helical manner as shown in Fig. 2. In the axial direction the spatial periodicity of the rotor magnets must equal that of the stator windings. If desired, the rotor permanent magnets may be replaced by windings fed with dc voltage from a brushless exciter and rectifier, as in wound rotor synchronous motors. Appropriate control equipment needs to be installed [23,24] such that the stator can produce any desired torque angle, and change this angle as quickly and smoothly as possible.

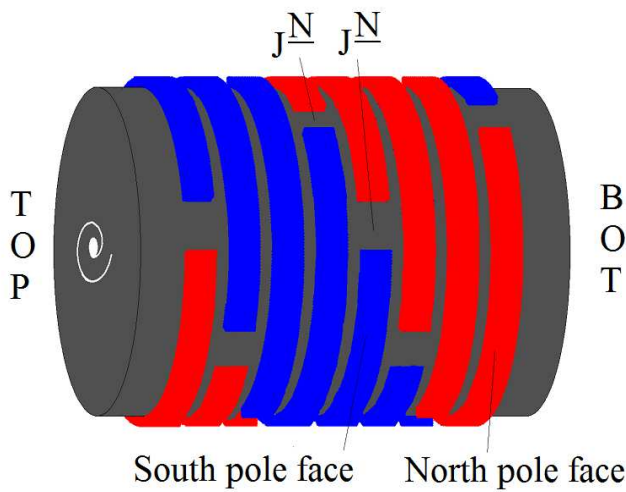

Figure 2 : *The rotor. Like the stator, it is dipolar. Red denotes a pole piece which is North on the outside while blue denotes one which is South on the outside. In a conventional motor, all the pieces would have been vertically aligned – here the location of the junction $J^{\underline{N}}$ from South to North rotates by 45° counterclockwise (CCW) as one moves up from one layer to the next, thus completing a full turn from the bottom to the top. Like the stator, this arrangement is also a right handed helix.*

The secondary windings on the stator side consist of two current wires close to the inner periphery, parallel to the cylinder axis and running the height of the stator. The two wires must be diametrically across from each other. On the rotor side, there are two similar wires, close to the outer periphery, diametrically across, parallel to the axis and running the height. All four wires are supplied with dc current, running in the same direction in each wire (say, from top to bottom). This current must be equal for the two stator wires and again for the two rotor wires. It is sufficient for the current magnitude to be constant in time. This arrangement is shown in Fig. 3. The return paths for these currents must be far away from the rotor-stator interface – near the centre in the case of the rotor and at a large radius in the case of the stator. In a fully permanent magnet-based construction, the dc wires can be replaced by a configuration of magnetic pole pieces with similar performance characteristics (the requirements are described in Subsection 3.1).

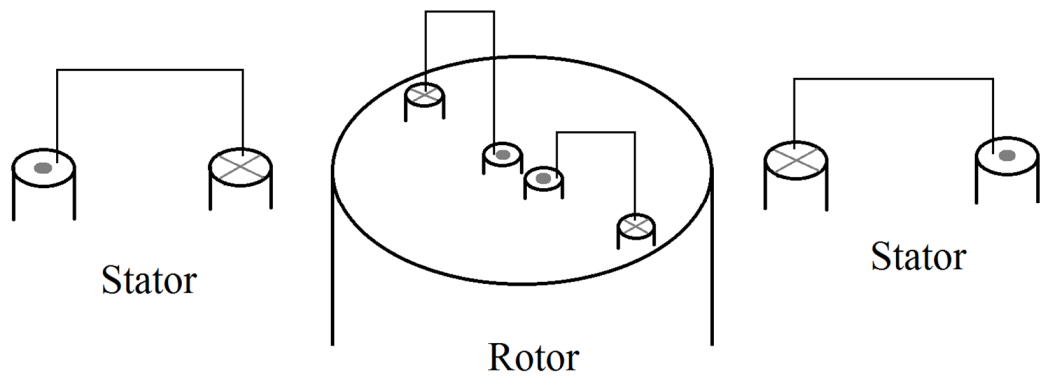

Figure 3 : *The secondary windings. Only the top part of the windings have been shown as they are invariant along the height. A cross denotes a current running in one direction, a dot the reverse.*

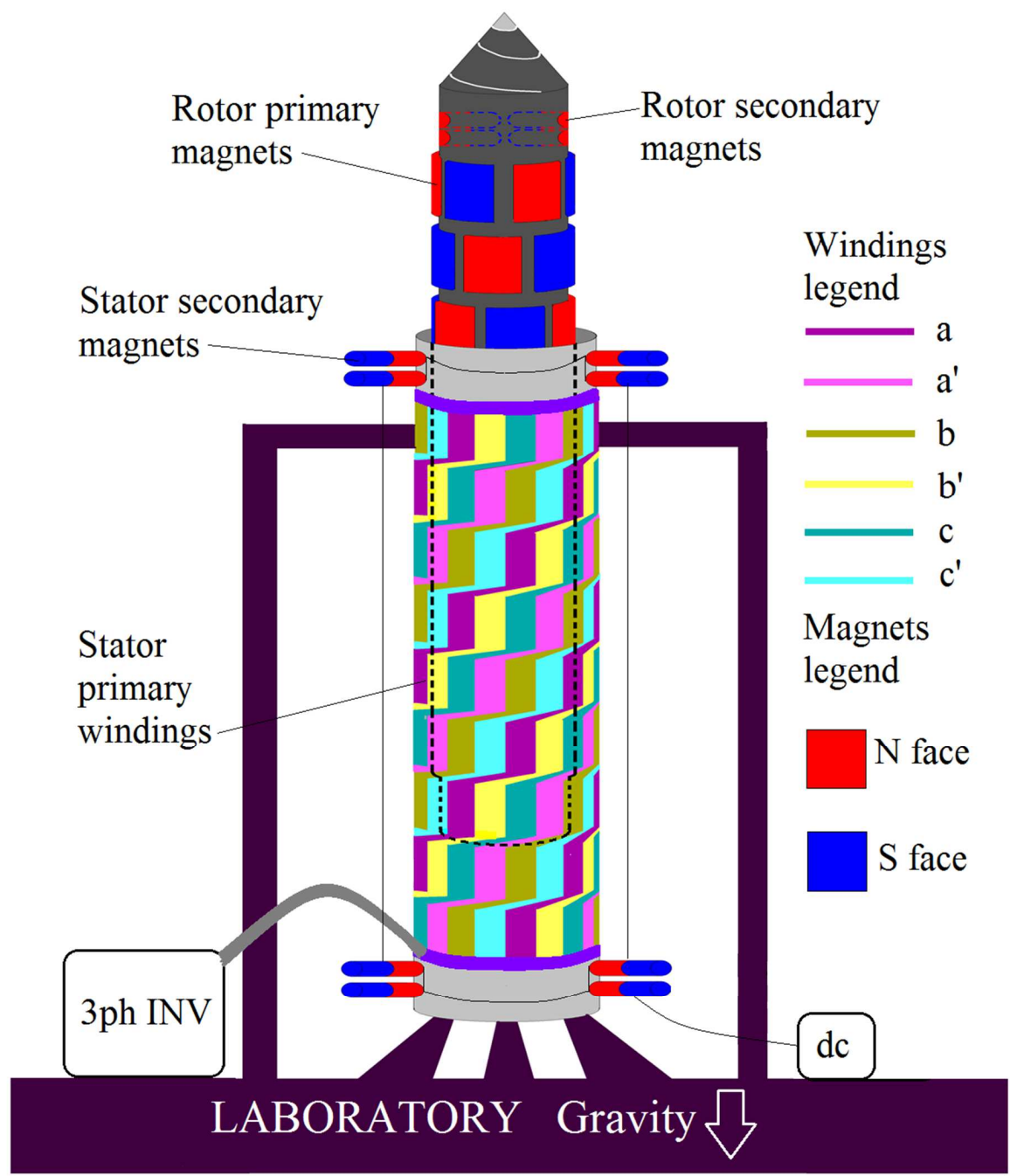

Figure 4 : *The total apparatus. The stator and rotor are similar to the ones of Figs. 1 and 2. This time however the polarity has been shown as 6, which is found later (Subsection 3.3) to be an optimum between performance and stability. The stator windings have been shown in magenta (phase a), yellow (phase b) and cyan (phase c) – a dark shade (1/3 less luminescence) indicates the proper branch while a bright shade (1/3 more luminescence) indicates the primed branch. Going around the stator at any fixed height, three copies of the basic unit a-b'-c-a'-b-c' can be seen. It is also noteworthy that the windings complete one full twist from the bottom to the top of the stator, even if the polarity suggests three twists – there are 9 layers of windings and they turn CCW 40° from each layer to the next. These windings are supplied by the three phase inverter shown. The rotor magnets also consist of three pairs of pole pieces in each layer. Again they are arranged in a staggered manner with 9 layers and 40° CCW twist from one to the next. For diagrammatic clarity, the rotor has been partially extruded from the stator cavity and is visible in silhouette in the remaining portion. The decoration on the spinner has been adapted from Rolls Royce aircraft engines. An implementation of the secondary winding in terms of magnets has been shown here. These are mounted on the top and bottom of the stator and rotor with all the North pole pieces (or all South) near each other and the opposite poles far away. This setup will produce qualitatively similar effects as the windings of Fig. 3 (which remains the configuration to be analysed in this Article). The magnets on the stator are electromagnets, shown supplied with a dc source. The rotor magnets are permanent magnets.*

A schematic view of the entire apparatus is shown in Fig. 4. This completes the design of the bearing, and hence this Subsection.

### 1.2 The Physics behind the bearing

The aim of this Subsection is to motivate the hypothesis that the apparatus described above will support a heavy rotor and exhibit a stable confined state. The starting point for this is the conventional PMSM. Since most motors are considerably long relative to the radius, it is assumed that the systems under study in this Article extend infinitely in the axial direction. In a conventional motor, this assumption reduces the electromagnetism to two dimensions; in the bearing treated here, it allows end effects to be rejected from consideration. Further, since the thicknesses of the stator and rotor are negligible compared to their radii, the currents flowing inside them can be treated as surface currents i.e. currents per unit length, which are traditionally denoted by **K** [25]. Henceforth, the word 'surface' will be understood to be implicit when dealing with currents in the motor and the bearing. In the conventional motor, the stator current is a function of the azimuthal angle in cylindrical coordinates $\rho,\theta,z$. Since it is periodic it can be expanded in a Fourier series where the order of the fundamental harmonic is half of the polarity of the motor. Since multiple harmonics give rise to ripple torques, stators and rotors are designed to be as close to single-harmonic as possible. Hence, with negligible error the stator current can be written as

$$\begin{aligned}\mathbf{K}_s &= \hat{\mathbf{z}}\left(K_{sd}\cos n\theta + K_{sq}\sin n\theta\right) \\ &= \hat{\mathbf{z}}K_s\cos n(\theta-\chi)\end{aligned} \quad , \tag{1.1}$$

where subscript $s$ denotes stator, $2n$ is the motor polarity, $d$ and $q$ refer to the direct and quadrature components of current and $\chi$ (whose range is from 0 to $2\pi/n$) is the angular position of the maximum of the current distribution. $K_{sd}$ and $K_{sq}$, or equivalently $K_s$ and $\chi$, are functions of time, regulated by the inverter. In most cases, they are chosen so that in the steady state of operation a rotating magnetic field of fixed strength is set up inside the stator cavity.

On the rotor, permanent magnets are mounted which again create an effective fundamental harmonic current

$$\begin{aligned}\mathbf{K}_r &= \hat{\mathbf{z}}\left(K_{rd}\cos n\theta + K_{rq}\sin n\theta\right) \\ &= \hat{\mathbf{z}}K_r\cos n(\theta-\nu)\end{aligned} \quad . \tag{1.2}$$

The rotor and stator are by definition coaxial and concentric. The interaction between the current in the rotor magnets and the magnetic field at the rotor surface causes a force and hence a resultant torque on the rotor. Since the rotor's own magnetic field cannot give rise to a torque on itself, the relevant interaction is actually between $\mathbf{K}_r$ and the stator magnetic field $\mathbf{B}_s$. Hence, the $\mathbf{B}_s$ produced everywhere by (1.1) must be calculated and then the product of $\mathbf{K}_r$ and $\mathbf{B}_s$ integrated all over the rotor surface to find the torque. The result for this is

$$\begin{aligned}T_z &\propto K_rK_s\sin n(\chi-\nu) \\ &= K_rK_s\sin\delta\end{aligned} \quad , \tag{1.3}$$

where the constant of proportionality is positive and $\delta$ is called the torque angle [26]. Note that $\delta$ is the angle from the maximum of the rotor current to the maximum of the stator current distribution, measured parallel to the direction of rotation. In a PMSM, the electromagnetic forces are independent of the rotor speed as the strengths of the stator and rotor currents remain constant at all speeds. A schematic representation of a conventional 2-pole ($n$=1) motor is shown in Fig. 5L.

In this Figure the sinusoidal current distributions on stator and rotor are represented by the annuli of variable thickness. A 'x' denotes current flowing in +$z$ direction i.e. out of the page while an 'o' denotes current flowing in –$z$ direction. Calculations show that the magnetic field due to (1.1) with $n$=1 is uniform inside the stator. This field goes from left to right and is depicted using the black lines. It should now be crossed with the rotor currents to obtain the force, which is thus directed perpendicular to both the field and the current. It is shown in the Figure by the green arrows. The forces lead to the torque on the rotor. The sin$\delta$ dependence of torque is intuitively clear – if $\delta$=90º the forces are all tangential and produce maximum torque while if $\delta$=0 the forces are purely radial and produce no torque at all.

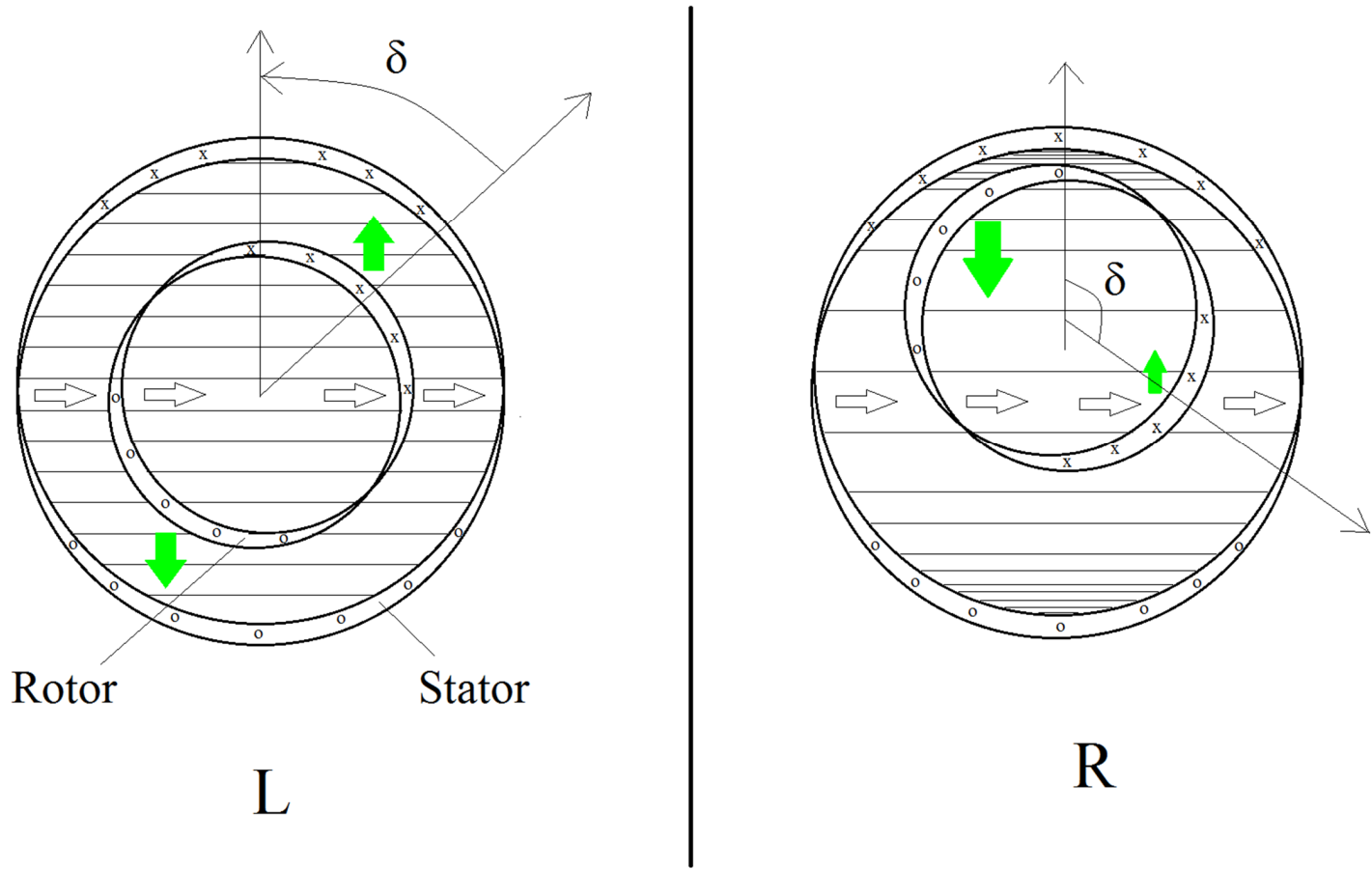

Figure 5 : *Left panel (L) shows cross-sectional view of a conventional motor while right panel (R) shows a two-dimensional magnetic bearing. The thin black lines are the stator magnetic field which goes from left to right. In (L) the field is uniform while in (R) a gradient has been imposed.*

The next question concerns the nature of the force if the rotor is displaced from its position at the centre of the stator. This is the situation encountered inside the magnetic bearing; for the apparatus to be stable the force must oppose the displacement. For the configuration of Fig. 5L, the answer is apparent from inspection – since the stator magnetic field is uniform, there will be no force on the rotor when it is displaced. A fundamental requirement essential to the production of a force is a spatial inhomogeneity in the magnetic field. Indeed, the uniform field is a speciality of the $n$=1 configuration; for all other polarities the field strength increases from the centre of the stator to the periphery. Because those fields are difficult to visualize however, a schematic Figure is prepared wherein the polarity of two is retained and a spatial gradient inserted by hand, in a manner which does not violate $\nabla \cdot \mathbf{B} = 0$. This modified field is seen in Fig. 5R – there is a gradient in **B** perpendicular to the field lines. The rotor is now displaced in the direction of the imposed gradient, as portrayed in the Figure. Clearly, a second quadrant $\delta$ as shown will give rise to a restoring force i.e. a force opposite to the displacement. This is desirable as it implies potential stability. Conversely, a first quadrant $\delta$ will cause the rotor to be displaced further from its equilibrium position, which is a sure sign of instability.

The two-dimensional nature of this configuration obviates the possibility of its generating a force in the *z* direction. To achieve magnetic levitation however a lift is clearly required, and this is where the helicity enters the picture. The primary difference between the conventional windings and helical windings shown in Fig. 1 ff. is that the current in the former is entirely axial whereas that in the latter has both axial and tangential components. In the conventional PMSM, the axial rotor current interacts with the radial component of stator field to produce a tangential force. In the helical scenario, if the helicity is small, the stator field can be assumed to be similar to that of the conventional case. However, the rotor current has an extra tangential component, so the radial field will interact with this current to produce an axial force. The size of this force will be of the same order as that of the rotor torque (divided by the rotor radius), and the force and torque will be proportional to each other. This axial force can be used to support the weight of the rotor. To obtain the direction of this force, it is assumed that the windings are helix dexter as shown in the Figures – the wires turn counter-clockwise (CCW) as they rise. The magnetic force rule now implies that if the torque is clockwise (CW), the axial force will be directed upward, as is desired in the present application.

On the basis of Fig. 5 it was argued above that for a conventional PMSM bearing, a second quadrant torque angle would result in confinement of the rotor in the horizontal plane. Again, for small helicity, the fields would not be significantly different from the conventional case and that conclusion would presumably not be overturned. To determine the stability in the newly introduced axial direction, a situation is considered wherein the rotor is rotating CW with the axial force balancing gravity, when it is subject to a small upward displacement. Since the helix is right-handed, this is equivalent to a small CCW rotation. Since $\delta$ is in the second quadrant (recall that it must be measured CW from the rotor to the stator), a CCW rotation of the rotor causes its value to increase, and the torque weakens. Then since force and torque are proportional, there is a commensurate reduction in the axial force. Finally since the axial force is upward, the resultant force on the rotor is now downwards i.e. opposite to the displacement and the rotor thus remains axially confined.

The argument about the restoring forces in the $x$ and $y$ directions implies a structure for these modes of motion like

$$\begin{bmatrix} \ddot{x} \\ \ddot{y} \end{bmatrix} = -\begin{bmatrix} \Omega^2 & 0 \\ 0 & \Omega^2 \end{bmatrix} \begin{bmatrix} x \\ y \end{bmatrix} , \qquad (1.4)$$

i.e. the effective spring constants in both $x$ and $y$ directions are the same. Such a structure is fragile because even small nonzero terms on the off-diagonal elements can compromise the stability. Indeed, the accurate calculation (Subsection 2.5) in fact yields that the two ciphers in the matrix above are replaced by small terms of opposite sign, which causes the system normal mode frequencies to acquire small imaginary parts, resulting in a loss of stability. The only way to restore stability is to make the two $\Omega$'s unequal, which is done through the introduction of the secondary windings. This makes these windings an important component of the whole system, so attention is now devoted to their mechanism of operation.

To study the effect of the secondary windings, the configuration of Fig. 3 is considered, and it is assumed that the rotor is spinning rapidly, initially positioned at the stator centre. The strength of the electromagnetic interaction is appreciable only between the conductors on the stator internal and rotor external periphery; this interaction is repulsive. Further, since the force varies as the inverse of the separation, the most significant repulsion occurs only when the pairs of wires are the most proximal. If the rotor is displaced in the $x$ direction as per the convention defined in Fig. 6, at and near the points of closest approach, it will be closer to the right stator wire and further away from the left stator wire. This will result in a force in the $-x$ direction, thus the rotor is expected to show harmonic oscillator behaviour in this direction. On the other hand, if the rotor is displaced in the $y$ direction, at the point of closest approach it will be 'above' both the stator wires and this will tend to push it further away. Thus the rotor is expected to show a harmonic repeller behaviour in this direction. In consequence, the secondary windings will cause the $x$ and $y$ spring constants of the combined system to become unequal.

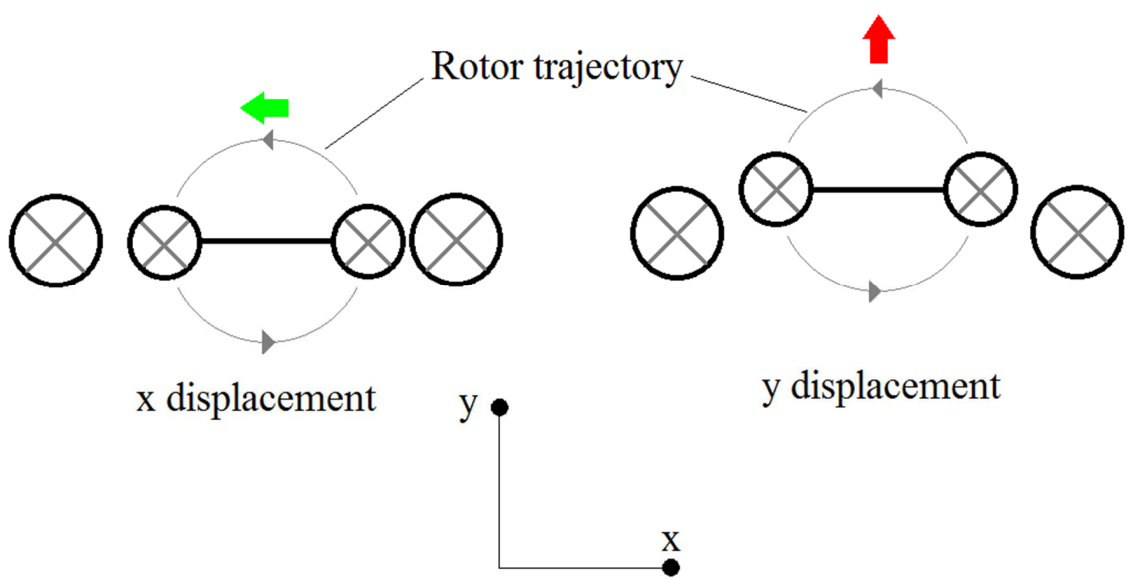

Figure 6 : *Cross-sectional view of the interacting currents from the secondary windings of Fig. 3. Left and right panels show the configurations of closest approach between stator and rotor wires in response to displacements in the x and y direction; the arrows indicate the resulting forces.*

These arguments motivate the effectivity of the proposed structure. Two issues should be noted. One is that the stability of rotational modes has not been found through these arguments, but will be obtained as a consequence of the calculation which follows. The second is that the existence of a confined state does NOT amount to a violation of Earnshaw's theorem. This is because the constant second quadrant $\delta$ is a fundamentally time-dependent step and the theorem does not hold for time-dependent magnetic fields. The argument might be made that, during steady state operation, the magnetic fields appear static in the synchronously rotating frame. However it is fallacious because the steady state is achieved through a balance of torque and load and Earnshaw's theorem is invalid in the presence of load. If the stator field were really held stationary, the rotor would work its way into a first quadrant $\delta$ and then lose stability.

The remainder of this Article is dedicated to quantitative proofs of the above arguments. The effect of the principal windings will be obtained in Section 2. The effect of the secondary windings, and hence the cumulative system stability, will be considered in Section 3. In that Section, the stiffness matrix will be presented and interpreted in terms of design parameters.

<div style="text-align:center">✻</div>

## 2 Linearized Equations for the Principal Windings Alone

In this Section, an expression is first obtained for the magnetic field due to the helical windings of Fig. 1. It is then used to derive the performance characteristics and dynamic model of a motor with helical windings. Subsequently the bearing is introduced and its mechanical equation derived. Finally the forces and torques on the rotor are calculated.

### 2.1 Helical motor current geometry

A helical motor is defined to be one where the stator and rotor windings are helical but the rotor is constrained to its default position concentric and coaxial with the stator. The forms of the stator and rotor currents for the helical PMSM are written by appropriate extension of the definitions (1.1) and (1.2) for the conventional PMSM. The equation for a single helically wound wire can be written as $\theta = kz + \varphi$ where $k$ is the wavenumber associated with the periodicity in the $z$ direction (i.e. $k = 2\pi/L$ where $L$ is the wavelength) and $\varphi$ runs from 0 to $2\pi$. A particular value of $\varphi$ uniquely identifies a particular wire. The current distribution in the cylinder is completely specified if the current flowing through each wire is prescribed, i.e. $K = f(\varphi) = f(\theta - kz)$ for some function $f$. For an alternating current motor which is being considered here, $f$ will also be a function of time. Due to slow variation of $f$ with respect to the speed of light, retardation can be ignored and the magnetostatic approximation invoked with negligible error. The direction of the current is parallel to the unit vector along each wire, which is $\hat{\boldsymbol{\theta}} \sin\alpha + \hat{\mathbf{z}} \cos\alpha$ where $\alpha = \arctan kR$ is the pitch angle of the helix and $R$ is the stator radius. Thus, the current in the winding can be expressed as

$$\mathbf{K} = f(\theta - kz)\left(\hat{\boldsymbol{\theta}} \sin\alpha + \hat{\mathbf{z}} \cos\alpha\right) \quad , \tag{2.1}$$

which is readily seen to be divergenceless for any choice of $f$.

Since $f$ is periodic with period $2\pi$, it can be expanded in a Fourier series, and the fundamental harmonic current written in terms of direct and quadrature components or equivalently in the magnitude-angle representation :

$$\begin{aligned}\mathbf{K}_s &= \left[K_{sd} \cos n(\theta - kz) + K_{sq} \sin n(\theta - kz)\right]\left[\hat{\boldsymbol{\theta}} \sin\alpha + \hat{\mathbf{z}} \cos\alpha\right] \\ &= \left[K_s \cos n(\theta - kz - \chi)\right]\left[\hat{\boldsymbol{\theta}} \sin\alpha + \hat{\mathbf{z}} \cos\alpha\right]\end{aligned} \quad , \tag{2.2}$$

and

$$\begin{aligned}\mathbf{K}_r &= \left[K_{rd} \cos n(\theta - kz) + K_{rq} \sin n(\theta - kz)\right]\left[\hat{\boldsymbol{\theta}} \sin\alpha + \hat{\mathbf{z}} \cos\alpha\right] \\ &= \left[K_r \cos n(\theta - kz - \nu)\right]\left[\hat{\boldsymbol{\theta}} \sin\alpha + \hat{\mathbf{z}} \cos\alpha\right]\end{aligned} \quad . \tag{2.3}$$

For the present purposes it is sufficient to find the magnetic field of (2.2) in the special case $\chi = 0$ since the generalization can be performed without difficulty.

### 2.2 The Magnetic field at the stator surface

It is seen that an eigenfunction expansion approach to the problem becomes unwieldy in practice due to the presence of an infinite number of terms in the series. Therefore the Biot Savart intergral approach is attempted in this case [27]. The Biot Savart law says that the magnetic field at position $\mathbf{r}$ is obtained in terms of wire currents $i$ at positions $\mathbf{r}'$ as $\mathbf{B}(\mathbf{r}) = \frac{\mu}{4\pi} \int \frac{i d\mathbf{l} \times (\mathbf{r} - \mathbf{r}')}{|\mathbf{r} - \mathbf{r}'|^3}$, and it is noteworthy that this formula is exact. The details of this approach are given in the Supplementary Material; the conclusion is that at a point on or near the surface of the cylinder, the dominant contribution will arise from the currents in the immediate vicinity and not from far away. This is of course a physically meaningful result – an analog of this is that near the Earth's surface, its gravitational field is like that of a sheet (i.e. uniform) and not like that of the entire sphere ($1/r^2$).

The foregoing conclusion motivates a transformation from the cylinder to a planar sheet which carries a current at an angle to its basic axes. At a point very close to either configuration, the currents have similar appearance and the magnetic fields should behave likewise. Performing the transformation, $\rho$ becomes the $\tilde{x}$ coordinate of the sheet, $\theta$ becomes the $\tilde{y}$ coordinate and $z$ the $\tilde{z}$ coordinate (the ~ notation on the coordinates serves to distinguish them from Cartesian coordinates $x,y,z$ associated with the cylinder itself). Since the cylinder is periodic in $\theta$, the sheet is made infinite in the $\tilde{y}$ direction and periodicity imposed with a characteristic wavenumber $\lambda$. The current in the sheet can be written as

$$\mathbf{K} = K \cos n\lambda(\tilde{y} - v\tilde{z})\left[\hat{\tilde{\mathbf{y}}}\sin\alpha + \hat{\tilde{\mathbf{z}}}\cos\alpha\right] \,, \tag{2.4}$$

where $v$ is a dimensionless number and the pitch $\alpha = \arctan v$. Using the divergenceless vector potential and separating variables, the following vector potential $\mathbf{A}$ and magnetic field $\mathbf{B}$ are obtained :

$$\mathbf{A}_{bot} = \mu K_s \tilde{l} \exp\frac{\tilde{x}}{2\tilde{l}} \cos\left(n\lambda(\tilde{y} - v\tilde{z})\right)\left[\hat{\tilde{\mathbf{y}}}\sin\alpha + \hat{\tilde{\mathbf{z}}}\cos\alpha\right] \,, \tag{2.5a}$$

$$\mathbf{A}_{top} = \mu K_s \tilde{l} \exp\frac{-\tilde{x}}{2\tilde{l}} \cos\left(n\lambda(\tilde{y} - v\tilde{z})\right)\left[\hat{\tilde{\mathbf{y}}}\sin\alpha + \hat{\tilde{\mathbf{z}}}\cos\alpha\right] \,, \tag{2.5b}$$

$$\mathbf{B}_{bot} = \mu K_s \exp\frac{\tilde{x}}{2\tilde{l}} \begin{pmatrix} \hat{\tilde{\mathbf{x}}}\left[n\lambda\tilde{l}(-\cos\alpha - v\sin\alpha)\sin(n\lambda(\tilde{y}-v\tilde{z}))\right] + \\ \hat{\tilde{\mathbf{y}}}\left[-\frac{1}{2}\cos\alpha\cos(n\lambda(\tilde{y}-v\tilde{z}))\right] + \hat{\tilde{\mathbf{z}}}\left[\frac{1}{2}\sin\alpha\cos(n\lambda(\tilde{y}-v\tilde{z}))\right] \end{pmatrix} \,, \tag{2.5c}$$

$$\mathbf{B}_{mid} = \mu K_s \exp\frac{-\tilde{x}}{2\tilde{l}} \begin{pmatrix} \hat{\tilde{\mathbf{x}}}\left[n\lambda\tilde{l}(-\cos\alpha - v\sin\alpha)\sin(n\lambda(\tilde{y}-v\tilde{z}))\right] + \\ \hat{\tilde{\mathbf{y}}}\left[\frac{1}{2}\cos\alpha\cos(n\lambda(\tilde{y}-v\tilde{z}))\right] + \hat{\tilde{\mathbf{z}}}\left[-\frac{1}{2}\sin\alpha\cos(n\lambda(\tilde{y}-v\tilde{z}))\right] \end{pmatrix} \,, \tag{2.5d}$$

where the characteristic length scale is defined as

$$\tilde{l} = \frac{1}{2n\lambda(1+v^2)^{1/2}} \,. \tag{2.6}$$

The inverse transformation is now effected i.e. features of this sheet solution are transferred to the helix. The following qualitative features are apparent from the sheet solution :

- At the surface of the sheet, $\mathbf{A}$ is parallel to $\mathbf{K}$ and the two waves are in completely phase
- The component of $\mathbf{B}$ normal to the sheet is completely out of phase with $\mathbf{A}$ and $\mathbf{K}$
- The component of $\mathbf{B}$ parallel to the sheet is completely in phase with $\mathbf{A}$ and $\mathbf{K}$ and is directed perpendicular to either of them

A trial vector potential which satisfies these conditions on the cylinder is

$$\mathbf{A}_{in} = \mu K_0 f_1(\rho)\cos n_0(\theta - kz)\left[\hat{\boldsymbol{\theta}}\sin\alpha + \hat{\mathbf{z}}\cos\alpha\right] \,, \tag{2.7a}$$

$$\mathbf{A}_{out} = \mu K_0 f_2(\rho)\cos n_0(\theta - kz)\left[\hat{\boldsymbol{\theta}}\sin\alpha + \hat{\mathbf{z}}\cos\alpha\right] \,, \tag{2.7b}$$

where $f_1$ and $f_2$ are two unknown functions. However, only their values and their derivatives at $\rho=R$ are required to find $\mathbf{B}$ exactly at the surface of the helix. These are found from the boundary conditions and in terms of the characteristic length scale

$$l = \frac{R}{2\left(\frac{n^2}{\cos^2\alpha} - \sin^2\alpha\right)^{1/2}} \,, \tag{2.8}$$

the field is obtained as

$$\mathbf{A} = \mu K_s l \cos n(\theta - kz)\left[\hat{\boldsymbol{\theta}}\sin\alpha + \hat{\mathbf{z}}\cos\alpha\right] \,, \tag{2.9a}$$

$$\mathbf{B}_{in} = \mu K_s \left| \hat{\boldsymbol{\rho}}\left[-\frac{nl}{R\cos\alpha}\sin n(\theta-kz)\right] + \hat{\boldsymbol{\theta}}\left[-\frac{1}{2}\cos\alpha\cos n(\theta-kz)\right] + \hat{\mathbf{z}}\left[\frac{1}{2}\sin\alpha\left(1+\frac{2l}{R}\right)\cos n(\theta-kz)\right] \right| \,, \tag{2.9b}$$

$$\mathbf{B}_{out} = \mu K_s \left| \hat{\boldsymbol{\rho}}\left[-\frac{nl}{R\cos\alpha}\sin n(\theta-kz)\right] + \hat{\boldsymbol{\theta}}\left[\frac{1}{2}\cos\alpha\cos n(\theta-kz)\right] + \hat{\mathbf{z}}\left[-\frac{1}{2}\sin\alpha\left(1-\frac{2l}{R}\right)\cos n(\theta-kz)\right] \right| \,. \tag{2.9c}$$

The magnetic field at the stator surface has thus been determined.

This expression enables the torque and axial force of a helical motor to be computed. It is recalled that the rotor torque is produced by the interaction of the rotor currents with the stator magnetic field; it can be written as

$$\mathbf{T} = \int_{-h}^{h} dz \int_{0}^{2\pi} d\theta \, r\mathbf{p} \times (\mathbf{K}_r \times \mathbf{B}_s) \quad , \tag{2.10}$$

where $2h$ is the height of the motor and $\mathbf{p}$ the position vector to any point on the rotor surface. Evaluating the product, and letting $r$ denote the rotor radius,

$$\begin{aligned} T_z &= \mu n r h l \left( K_{rd} K_{sq} - K_{sd} K_{rq} \right) \\ &= \mu n r h l K_r K_s \sin(\chi - \nu) = \mu n r h l K_r K_s \sin \delta \end{aligned} \tag{2.11}$$

These expressions have the same forms as those for conventional motors. The axial force can be obtained from an analogous calculation :

$$\begin{aligned} F_z &= -\mu n h l \tan \alpha \left( K_{rd} K_{sq} - K_{sd} K_{rq} \right) \\ &= -\mu n h l \tan \alpha K_r K_s \sin \delta \end{aligned} \tag{2.12}$$

It is noteworthy that the above two expressions are concurrent with the hypothesis made on physical grounds that the axial force is of the same size as the torque and that a CW torque produces a force directed upwards. Moreover, if there is no helicity then $\alpha=0$ and the axial force vanishes as it should. It must also be kept in mind that, insofar as $\mathbf{B}_s$ has been obtained only at the stator surface, the assumption of $r=R$ is implicit in the preceding equations.

## 2.3 The Mechanical equations of the bearing rotor

The rotor orientation relative to the stator frame is specified in terms of the three Euler angles $\varphi, \tilde{\theta}, \psi$ for precession, nutation and spin respectively. (The ~ on $\theta$ serves to avoid overlap with the stator cylindrical coordinates $\rho, \theta, z$ and will be removed once these coordinates have ceased to be essential to the calculation. These two applications of $\theta$ are both so thoroughly ingrained into today's vocabulary that an initial attempt at using $\beta$ to denote either angle was found to have an effect contrary to the desired intention.) The lab axes $x,y,z$ are chosen with the origin at the geometrical centre of the stator and $z$ along its axis of symmetry. The centre of mass (CM) of the rotor is displaced from the origin by $x_{CM}$, $y_{CM}$ and $z_{CM}$ to form the translated but parallel basis $X,Y,Z$. From this point the Eulerian precession $\varphi$ about the $z$-axis followed by Eulerian nutation $\tilde{\theta}$ about the intermediate $x$-axis lead to the rotor $a,b,c$ frame. Eulerian spin $\psi$ about the new $z$-axis leads further to the rotor $u,v,w$ frame, but the assumption of symmetry of the rotor i.e. $I_a=I_b=I$ ensures that the $a,b,c$ frame is sufficient for all calculations. The $a,b,c$ representation of the rotor angular velocity $\boldsymbol{\omega}$ is

$$\boldsymbol{\omega}_{abc} = \hat{\mathbf{a}}\dot{\tilde{\theta}} + \hat{\mathbf{b}}\dot{\varphi}\sin\tilde{\theta} + \hat{\mathbf{c}}\left(\dot{\psi} + \dot{\varphi}\cos\tilde{\theta}\right) \quad , \tag{2.13}$$

and the angular velocity of the $a,b,c$ frame itself is $\boldsymbol{\omega}^f = \boldsymbol{\omega} - \dot{\psi}\hat{\mathbf{c}}$. The translational equations are obvious while the rotational equations are obtained by writing Euler's equation in the $a,b,c$ frame i.e. $d\mathbf{L}/dt\big|_{\text{fixed}} = \dot{\mathbf{L}} + \boldsymbol{\omega}^f \times \mathbf{L}$, and equating the rate of change of $\mathbf{L}$ to the torque $\mathbf{T}$. Then, the rotor mechanical equations are

$$m\ddot{x}_{CM} = F_x \quad , \tag{2.14a}$$

$$m\ddot{y}_{CM} = F_y \quad , \tag{2.14b}$$

$$m\ddot{z}_{CM} = F_z - mg \quad , \tag{2.14c}$$

$$I\dot{\omega}_a + I_c \omega_c \omega_b^f - I\omega_b \omega_c^f = T_a \quad , \tag{2.14d}$$

$$I\dot{\omega}_b + I_a \omega_a \omega_c^f - I_c \omega_c \omega_a^f = T_b \quad , \tag{2.14e}$$

$$I_c \dot{\omega}_c = T_c - \Gamma \quad , \tag{2.14f}$$

where $\mathbf{F}$ and $\mathbf{T}$ are entirely electromagnetic in origin and $\Gamma$ is the load torque.

At the linear level the above equation system simplifies considerably. The precession $\varphi$ becomes indistinguishable from the spin $\psi$ and the two merge into a single angle whose derivative is $\omega_c$. The separation of these modes of motion is achieved by the nutation, which, in the present application, is desired to be confined about a zero value. The term $\omega_b$ becomes a product of two small quantities and so (2.14e) becomes redundant. The nonlinear terms vanish from (2.14d), being of the second order of smallness. Equation (2.14f) is also superfluous because the speed of the rotor is controlled externally, varying as per the application requirements. Even if perturbations $x_{CM}$, $y_{CM}$, $z_{CM}$ and $\tilde{\theta}$ cause $\omega_c$ to drift while themselves dying down, the external controller can always correct $\omega_c$ with an appropriate acceleration. Hence this equation can also be eliminated and the linearized mechanical equation set is

$$m\ddot{x}_{CM} = F_x \quad , \tag{2.15a}$$

$$m\ddot{y}_{CM} = F_y \quad , \tag{2.15b}$$

$$m\ddot{z}_{CM} = F_z + mg \quad , \tag{2.15c}$$

$$I\ddot{\tilde{\theta}} = T_a \quad . \tag{2.15d}$$

It thus remains to compute the three components of force and one component of torque to obtain the equation of motion of the rotor.

### 2.4 The Magnetic field at the rotor surface

The electromagnetic force and torque on the rotor can be calculated from integrals like (2.10) which run over the rotor surface. As a first step towards this, the stator field (2.9) must be developed from the surface of the stator into its interior. To this end, a first order Taylor expansion is performed (rotor radius assumed very close to stator radius), yielding $\mathbf{B}_s$ in a cylindrical annulus of thickness $\xi$ with the stator as the external bounding surface. Assuming that the pitch angle $\alpha$ remains constant in the entire annulus, the following is found :

$$\mathbf{B}_s(R-\xi,\theta) = \mu K_s \left| \begin{array}{l} \hat{\boldsymbol{\rho}}\left[ n\left(1-\dfrac{\xi}{2}\right)\left(-\dfrac{1}{(R-\xi)\cos\alpha}\right)\sin n(\theta-kz-\chi)\right] + \hat{\boldsymbol{\theta}}\left[-\dfrac{1}{2}(1-\eta\xi)\cos\alpha\cos n(\theta-kz-\chi)\right] + \\ \hat{\mathbf{z}}\left[\dfrac{1}{2}\left(1-\eta\xi+\dfrac{1-\xi/2}{R-\xi}\right)\sin\alpha\cos n(\theta-kz-\chi)\right] \end{array} \right| \tag{2.16}$$

where $\eta$ denotes the gradient in $\mathbf{B}_s$. This is the spatial inhomogeneity which was earlier shown to be essential for confinement stability. The value of this quantity is not yet known but physical arguments yield that it is positive; an approximate value will be presented in Subsection 3.3.

From this point onwards, the following operations yield $\mathbf{B}_s$ at the rotor surface :

- The cylindrical polar coordinates in the above expression are converted to Cartesian coordinates $x,y,z$. Here, for simplicity, the substitution $n=1$ is used. This step is impermissible in a conventionally wound motor or equivalent bearing because of the uniformity of the dipolar field but the helicity adds a gradient even in this case. The case of higher polarity will be examined in Subsection 3.3.
- All points on the rotor surface are expressed in terms of the coordinates $x,y,z$.

These steps are considerably cumbersome and are shown in the Supplements. During these steps, the smallness of the perturbations $x_{CM}$, $y_{CM}$, $z_{CM}$ and $\theta$ is used to retain only first order terms in these quantities.

### 2.5 The Electromagnetic force and torque

Below, the procedure to perform the integrals for force and torque is indicated briefly. As an exemplar, $F_b$ is considered. The magnetic force rule yields

$$F_b = \int_{-h}^{h} dc \int_{0}^{2\pi} d\gamma \, r \left( K_{rc} B_{sa} - K_{ra} B_{sc} \right) \quad . \tag{2.17}$$

Likewise, the only relevant component of torque $T_a$ has a similar form – opening out the triple product (2.10) with appropriate symbols,

$$T_a = \int_{-h}^{h} dc \int_{0}^{2\pi} d\gamma \, r \left[ rS \left( K_{ra} B_{sb} - B_{sa} K_{rb} \right) + c \left( K_{ra} B_{sc} - B_{sa} K_{rc} \right) \right] \quad . \tag{2.18}$$

The second term in the above integrand is similar to $F_b$.

Each integration in the above brings its own set of simplifications. The one over $\gamma$ is considered first. The various terms in $\mathbf{K}_r$ and $\mathbf{B}_s$ are all trigonometric functions of $\gamma$; when these are multiplied together and integrated over $\gamma$ they will reduce to the averages of the products. These averages, of all trigonometric monomials upto and including the 6$^{\text{th}}$ degree, are summarized in the below table :

$$<C^2>=<S^2>=1/2$$
$$<C^4>=<S^4>=3/8 \quad <C^2 S^2>=1/8 \quad\quad\quad (2.19)$$
$$<C^6>=<S^6>=5/16 \quad <C^4 S^2>=<S^4 C^2>=1/16$$

and any average not listed here is zero. After averaging, like terms are clubbed together and simplified using standard trigonometric identities. This yields two types of terms – half of them are trigonometric functions of $2kc+\psi+\chi$ while the other half are trigonometric functions of the torque angle $\delta$. For the $c$ integration, terms of the latter type are considered first. When the integration is performed, the terms which do not feature $c$ anywhere else (in fact this is majority of the terms) get trivially multiplied by $2h$, while the other terms involve integration of simple monomials.

Attention is now drawn to the terms which have cos and sin of $2kc+\psi+\chi$. In a typical magnetic bearing application, the rotor spins fast; since the torque angle is more or less constant, $\chi$ has the same form as $\psi$. These terms thus constitute rapidly varying quantities, which can on the one hand cause vibrations of the rotor and on the other hand create weird effects on system stability, as in a Kapitsa pendulum. Their effect however will be greatly mitigated if the windings are such that $\int_{-h}^{h} \cos 2kc \, dc = \int_{-h}^{h} \sin 2kc \, dc = 0$, i.e. if the rotor and stator windings complete integer number of turns from the bottom to the top of the helix. If this relation is satisfied, all such terms which have no other dependence on $c$ will evaluate to identically zero – the ones which are not zero (integrals of $c\sin kc$, $c^2 \cos kc$ etc.) will also be quite small. Because of the smallness and fastness of these residual terms, they will be ignored in the remainder of this calculation.

When these steps are performed, the equations for force and torque in terms of $x_{CM}$, $y_{CM}$ and $\theta$ can be obtained. They are cumbersome and are shown in the Supplements. However, they can be easily deduced from the matrix (3.7) displayed in the next Section; the only extra term is an eccentricity-independent $F_z$ which reduces to (2.12) when $r=R$.

<div style="text-align:center">✱</div>

# 3 Full System Equations and Stability Analysis

In this last Section of this Article, the effect of the secondary windings is first analysed. Subsequently the stiffness matrix of the entire system is written down and a discussion of stability presented.

### 3.1 The Effect of the secondary winding

Inspection of the $x$ and $y$ components of force indicates that a second quadrant torque angle $\delta$ is indeed necessary if confinement is to be at all achieved and the rotor weight also balanced. Unfortunately, considering the $x$ and $y$ motions alone, the structure is

$$\frac{d^2}{dt^2}\begin{bmatrix} x_{CM} \\ y_{CM} \end{bmatrix} = -\begin{bmatrix} \Omega^2 & -\varepsilon^2 \\ \varepsilon^2 & \Omega^2 \end{bmatrix}\begin{bmatrix} x_{CM} \\ y_{CM} \end{bmatrix} \quad , \tag{3.1}$$

which (cf. Subsection 1.2) is unstable. Hence the secondary windings are now introduced and the assertion verified that they indeed constitute a harmonic oscillator in one direction and a harmonic repeller in the other. The two stator conductors each carrying current $i_s$ are assumed to be mounted at $x=R_1$ and $x=-R_1$; the two rotor conductors carrying current $i_r$ are opposite to each other at radius $r_1$ and the rotor makes an angle $\psi$ with the stator axes. The force per unit length between two parallel current carrying wires at locations $\mathbf{r}_1$ and $\mathbf{r}_2$ in the plane [25] is

$$\mathbf{f} = \frac{\mu i_1 i_2}{2\pi} \frac{\mathbf{r}_1 - \mathbf{r}_2}{|\mathbf{r}_1 - \mathbf{r}_2|^2} \quad . \tag{3.2}$$

Using this, the force on each rotor wire due to each stator wire can be readily calculated and the various contributions subsequently added. This yields cumbersome trigonometric functions of $\psi$. However, it is recognized that $\psi$ is a fast variable, whereby slowly varying quantities can be obtained by averaging over it. These integrations are performed numerically, yielding

$$F_x = x_{CM} \int_0^{2\pi} d\psi \begin{bmatrix} \dfrac{2}{R_1^2 + r_1^2 - 2R_1 r_1 \cos\psi} - 4\dfrac{(R_1 - r_1 \cos\psi)^2}{(R_1^2 + r_1^2 - 2R_1 r_1 \cos\psi)^2} + \\ \dfrac{2}{R_1^2 + r_1^2 + 2R_1 r_1 \cos\psi} - \dfrac{4(R_1 + r_1 \cos\psi)^2}{(R_1^2 + r_1^2 + 2R_1 r_1 \cos\psi)^2} \end{bmatrix} = -q_1 x_{CM} \quad , \tag{3.3a}$$

$$F_y = y_{CM} \int_0^{2\pi} d\psi \begin{bmatrix} \dfrac{2}{R_1^2 + r_1^2 - 2R_1 r_1 \cos\psi} - \dfrac{4 r_1^2 \sin^2\psi}{(R_1^2 + r_1^2 - 2R_1 r_1 \cos\psi)^2} + \\ \dfrac{2}{R_1^2 + r_1^2 + 2R_1 r_1 \cos\psi} - \dfrac{4 r_1^2 \sin^2\psi}{(R_1^2 + r_1^2 + 2R_1 r_1 \cos\psi)^2} \end{bmatrix} = q_2 y_{CM} \quad , \tag{3.3b}$$

which act as the definitions of $q_1$ and $q_2$. It is noteworthy that they are both positive. Of course, since the configuration is invariant in $z$, there is no $F_z$. The nutation angle does not change these expressions since the relevant current component is the cosine, which is of size unity for small nutations. Likewise, perturbations in $x$ and $y$ do not cause the rotor to nute. A small nutational tendency is however automatically amplified – a suitably approximate estimate of the nutational torque is the force between the conductors times $h\sin\theta$ times a numerical factor to allow for the averaging. This factor is assumed to be $1/2$, yielding

$$T_a = \frac{\mu i_r i_s h^2}{2\pi(R_1 - r_1)} \theta = q_3 h^2 \theta \quad . \tag{3.4}$$

The exact value of $q_3$ is of secondary relevance as the nutational torque is a small correction to the contribution of the principal windings, and there is no degenerate eigenvalue in this part of the system to make small terms important. It is also noted that the secondary windings will not interact with the principal windings on a slow time scale because their polarities are different and the forces and torques will average out to zero over each rotation of the rotor. This step completes the determination of forces and torques on the rotor and the stiffness matrix of the combined system can now be written down.

## 3.2 The Stiffness matrix

The equation of motion is written by combining the right hand side force and torque terms from the Supplement with the left hand side (2.15). Physical and mathematical clarity is gained if the variable $\theta$ is replaced by $w=r\theta$, which also has dimensions of length. Then, the equation of motion possesses the structure

$$\frac{d^2}{dt^2}\begin{bmatrix} x_{CM} \\ y_{CM} \\ z_{CM} \\ w \end{bmatrix} = -\kappa \begin{bmatrix} x_{CM} \\ y_{CM} \\ z_{CM} \\ w \end{bmatrix}, \quad (3.5)$$

where $\kappa$ denotes the stiffness matrix. This matrix can be economically represented with the substitution $\zeta = 2\mu r h K_r K_s$, the notations $C=\cos\alpha$ and $S=\sin\alpha$ and the following variable definitions [where the characteristic length from (2.8) is repeated] :

$$l = \frac{R}{2\left(\dfrac{n^2}{\cos^2\alpha} - \sin^2\alpha\right)^{1/2}}, \quad (3.6a)$$

$$k_1 = \frac{1}{R\cos\alpha}, \quad (3.6b)$$

$$k_2 = k_1 - \frac{l\cos\alpha}{R^2}, \quad (3.6c)$$

$$k_3 = \eta + \frac{1}{2R} - \frac{l}{R^2}, \quad (3.6d)$$

$$k_4 = \frac{1}{r}\left[-lk_1 + \frac{k_2}{4R}(R^2 - r^2)\right], \quad (3.6e)$$

$$k_5 = \frac{1}{r}\left[1 - \frac{\eta}{2R}(R^2 - r^2)\right], \quad (3.6f)$$

$$k_6 = \frac{1}{r}\left[1 + \frac{l}{r} - \frac{k_3}{2R}(R^2 - r^2)\right]. \quad (3.6g)$$

In terms of these the matrix is

$$\kappa = \begin{bmatrix} \dfrac{1}{m}\left[-\zeta\left(\dfrac{Ck_4}{4} + \dfrac{C^2 k_5}{8} + \dfrac{C^2\eta r}{8R} + \dfrac{S^2 k_3 r}{8R}\right)\cos\delta + q_1\right] & \dfrac{\zeta}{m}\left(\dfrac{Ck_2 r}{8R} - \dfrac{Ck_4}{4} + \dfrac{C^2 k_5}{8} - \dfrac{S^2 k_6}{8}\right)\sin\delta & 0 & \dfrac{\zeta}{m}\left(-\dfrac{Crkk_4}{2} - \dfrac{CSk_5}{8} - \dfrac{C^2 rkk_5}{8} + \dfrac{CSk_6}{2}\right)\cos\delta \\ \dfrac{\zeta}{m}\left(-\dfrac{Ck_2 r}{8R} + \dfrac{Ck_4}{4} - \dfrac{C^2 k_5}{8} + \dfrac{S^2 k_6}{8}\right)\sin\delta & \dfrac{1}{m}\left[-\zeta\left(\dfrac{Ck_4}{4} + \dfrac{C^2 k_5}{8} + \dfrac{C^2\eta r}{8R} + \dfrac{S^2 k_3 r}{8R}\right)\cos\delta - q_2\right] & 0 & \dfrac{\zeta}{m}\left(-\dfrac{Sk_5}{2} + \dfrac{CSk_5}{8} + \dfrac{C^2 rkk_5}{8} + \dfrac{S^2 rkk_6}{8}\right)\sin\delta \\ 0 & 0 & \dfrac{\zeta}{m}\left(\dfrac{Srkk_4}{2}\right)\cos\delta & 0 \\ \dfrac{\zeta}{I}\left(-\dfrac{5Sr^2 k_4}{16} - \dfrac{Sr^2 k_5}{8} + \dfrac{CSr^2 k_5}{16} + \dfrac{CS\eta r^3}{16R}\right)\cos\delta & \dfrac{\zeta}{I}\left(\dfrac{Sk_2 r^3}{8R} - \dfrac{CSr^2 k_5}{32} + \dfrac{Sr^2 k_5}{32}\right)\sin\delta & 0 & \dfrac{1}{I}\left[-\zeta\left(\dfrac{Sr^3 kk_4}{2} + \dfrac{Ch^2 k_4}{12} - \dfrac{S^2 r^2 k_5}{4} + \dfrac{Ch^2 k_5}{24} + \dfrac{C^2 h^2 \eta r}{24R} + \dfrac{S^2 h^2 k_3 r}{24R}\right)\cos\delta - q_3 h^2\right] \end{bmatrix}$$

(3.7)

The next subsection of this Article will be dedicated to a discussion of stability.

### 3.3 Analysis of system stability

As a first approximation, (3.5,7) can possibly describe stable motions if the stiffness matrix is positive definite, and by extension, its diagonal elements are strongly positive. This is not a sufficient condition for stability, as will soon be shown, but is indicative. It can be seen that the diagonal elements on the matrix are positive if $\cos\delta<0$ i.e. $\delta$ lies in the second quadrant. This is in accordance with our expectations. Notable also is that terms featuring the gradient $\eta$ appear directly on all the diagonal terms, hence the system stability can be augmented by increasing the value of this parameter. If $\alpha=0$ then $k=0$ and $\kappa_{33}$ is zero as expected (it may be recalled that a conventional motor is neutral to perturbations in $z$). As $\alpha$ is increased, $\kappa_{33}$ becomes non-zero; since $k_4$ is negative, $\kappa_{33}$ is positive for second quadrant $\delta$, again consistent with intuition.

Thus, $\eta$ and $\alpha$ clearly are control parameters which determine the system stability. In a practical arrangement however both of them cannot vary arbitrarily. The system size is determined from the application requirements; the arguments of Subsection 2.5 indicate that the windings must perform whole numbers of twists over the height, thus constraining $k$ and hence $\alpha$ to a set of discrete values. The design parameter with the most obviously pronounced dependence on $\eta$ is the motor polarity, which is again a discrete variable. $\delta$ too is a control parameter with limited variation – the triplicate conditions of balancing the load torque and the rotor weight while retaining stability will impose severe constraints on its range. The only unrestricted control parameter is the strength of current in the secondary windings $i_r$ and $i_s$, and the resultant $q_1$, $q_2$ and $q_3$.

The absence of damping terms in the linearized system (3.5,7) means that stability will be of the oscillatory type and not of the absolute type i.e. perturbations will continue to oscillate rather than decaying in time. The analysis of the system is non-trivial however because the stiffness matrix $\kappa$ is not symmetric. This means that the system is non-conservative, which indeed it is because of the presence of the external drive from the inverter. Discussions of such systems may be found in [28-32]. In particular, the work of P GALLINA [31] mentions the stability criterion for such systems explicitly (see the Supplement for a summary). The criterion requires the parameters

$$a_1 = \kappa_{11} + \kappa_{22} + \kappa_{44} \quad , \tag{3.8a}$$

$$a_2 = \kappa_{11}\kappa_{22} + \kappa_{11}\kappa_{44} + \kappa_{12}^2 - \kappa_{14}\kappa_{41} + \kappa_{22}\kappa_{44} - \kappa_{24}\kappa_{42} \quad , \tag{3.8b}$$

$$a_3 = \kappa_{11}\kappa_{22}\kappa_{44} - \kappa_{11}\kappa_{24}\kappa_{42} + \kappa_{12}^2\kappa_{44} - \kappa_{12}\kappa_{24}\kappa_{42} + \kappa_{12}\kappa_{24}\kappa_{41} - \kappa_{14}\kappa_{22}\kappa_{41} \quad , \tag{3.8c}$$

and in terms of these, can be written as

$$a_1, a_2, a_3 > 0 \quad , \tag{3.9a}$$

$$a_1^2 - 3a_2 > 0 \quad , \tag{3.9b}$$

$$a_1^2 a_2^2 - 4a_1^3 a_3 + 18a_1 a_2 a_3 - 4a_2^3 - 27a_3^2 > 0 \quad . \tag{3.9c}$$

This condition allows the determination of stability without the time-consuming reduction of (3.5,7) to a system of eight first order linear equations and then verification that all the real parts are exactly zero. Hence it will be used in this Article to evaluate the stability of the magnetic bearing.

Below, a series of stability charts will be presented as the polarity, pitch and secondary current are varied. The calculations for polarity will involve obtaining an approximate $\eta(n)$ from the conventional motor case (this will be an underestimate), substitution of this value into (3.6a) to recalculate the various $k$'s, and then substitution of everything into (3.7). The value of $\eta$ for a conventional motor is

$$\eta = \frac{\mu K_s (n-1)}{2R} \quad . \tag{3.10}$$

Now, numerical values are assigned to the various parameters. A household scale application such as a maglev spin dryer or a maglev ceiling fan is imagined, leading to the following values :

- $2R$=11 cm or, since SI Units must be used consistently in numerical work, 0.11 m
- $2r$=10 cm or 0.10 m
- $2h$=60 cm or 0.6 m (this allows a 30° pitch angle for the lowest mode)
- $K_r K_s$=1.2x$10^8$ A$^2$/m$^2$ ($K_r$ and $K_s$ will be varied while keeping their product i.e. torque, power etc constant)
- $i_r$=220 A
- $\mu$=$10^{-4}$ SI Units (about 100 times that of free space)
- $R_1$=6 cm or 0.06 m

- $r_1$=4.5 cm or 0.045 m
- $I/m \approx (2h)^2/12$=300 cm² or 0.03 m²

The torque angle $\delta$ is chosen fixed at 150° for all plots, which represents a balance between high performance and stability. For the first case, $K_s$ is taken as 6000 A/m while $K_r$ is taken as 20000 A/m. The pitch is taken as the lowest value common to all polarities i.e. one full turn over the entire height. The pitch angle is 30°. The lift force and torque are 0.96 kgf and 0.84 Nm for the quadrupolar case, and decrease strongly with increase in polarity. The polarity is varied in steps from $n$=2 to $n$=11. The secondary stator current is also varied, assuming the secondary rotor current to remain constant at 220 A. Figure 7 shows the stability diagram – a point is marked green if operation there is stable and red if unstable.

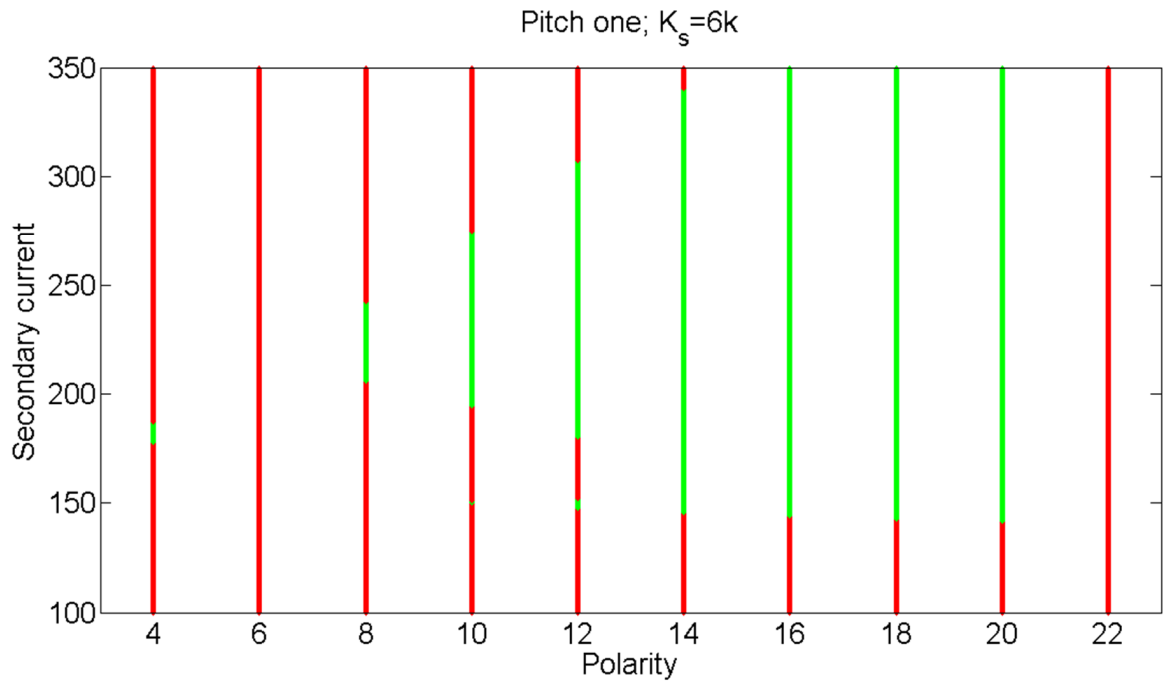

Figure 7 : *Stability diagram as polarity and secondary stator current are varied. Green shows a stable operating point, red an unstable one.*

The results are in agreement with expectations : the quadrupolar and sextupolar rotors are unstable (the tiny green island on the red line at $2n$=4 is likely a computational error) and thereafter the region of stability keeps increasing as the polarity and hence $\eta$ increases. The sudden disappearance of stability at polarity 22 is however unexpected and requires a discussion. The problem is seen to arise from the $\kappa_{33}$ term, which contains a factor of $k_4$. As the characteristic length $l$ keeps decreasing, $k_4$ from (3.6e) changes from negative to positive, causing $\kappa_{33}$ to change sign. Physically, a change in the sign of $k_4$ means, if it is traced back to $\mathbf{B}_s$ and expressed in cylindrical coordinates, that $B_{sp}$ changes sign at a certain radius, which is implausible. This indicates that the high polarity is outside the range of validity of the Taylor expansion from (2.9) to (2.16) – the expansion is valid in an annulus of thickness of order $l$, and when $l$ diminishes beyond a certain value, the expansion ceases to be plausibly accurate even at the rotor surface. In reality, the field does not reverse sign but does become extremely weak, which is why the lift and torque degrade rapidly as polarity increases. (Of course, the appropriate rotor mass has been used for all the calculations, and the moment of inertia obtained through multiplication by 300 cm², as mentioned above.) Thus, it can be concluded that the instability at $n$=11 is spurious but the design point is of negligible practical utility.

In the next figure, the pitch is increased to two full turns over the height, thus giving an angle of almost 50°. The lift force increases to 1.33 kgf and the torque decreases to 0.58 Nm for the quadrupole. The spurious instability mentioned above exists beyond $2n$=14 so those lines are shown in yellow. However, the only stable polarities are the high ones, where the lift force and torque are very small.

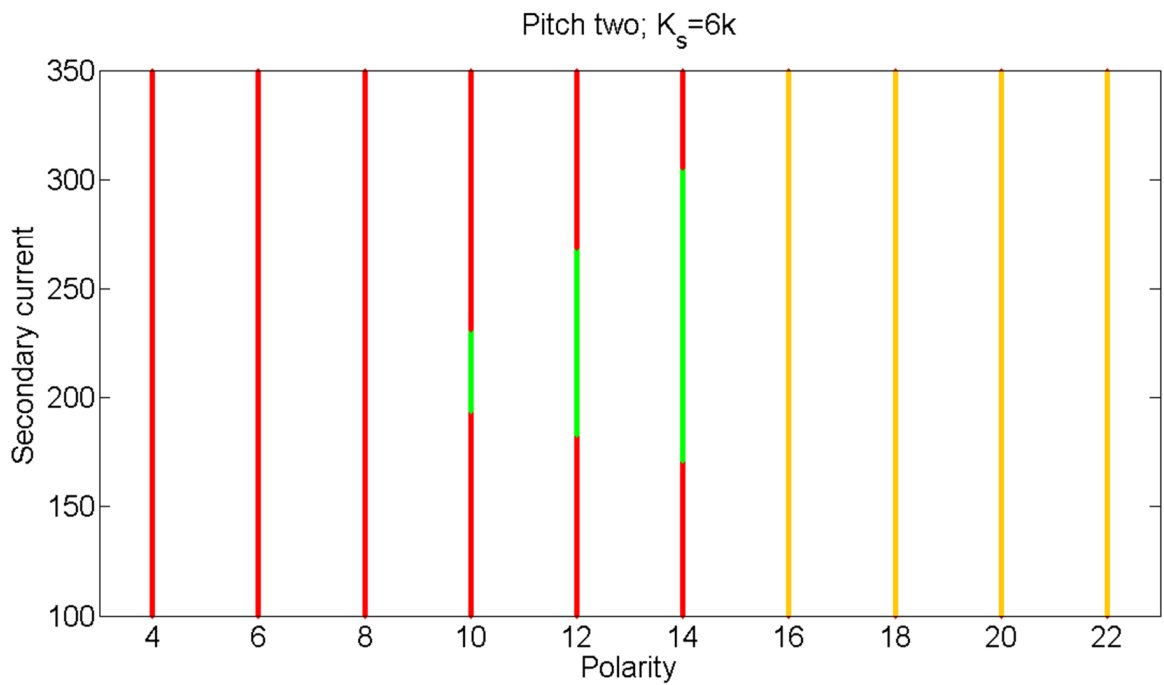

Figure 8 : *Stability diagram with different parameter values.*

The situation is saved by turning to factors affecting $\eta$ other than polarity. One such factor is $K_s$ and the motor parameters are accordingly revised to $K_s$=16000 A/m and $K_r$=7500 A/m. The pitch is again set to 30º; the lift force and torque for the sextupole are 0.56 kgf and 0.48 Nm while those for the octupole are 0.36 kgf and 0.31 Nm respectively. This time there are substantial stability zones in the viable polarities.

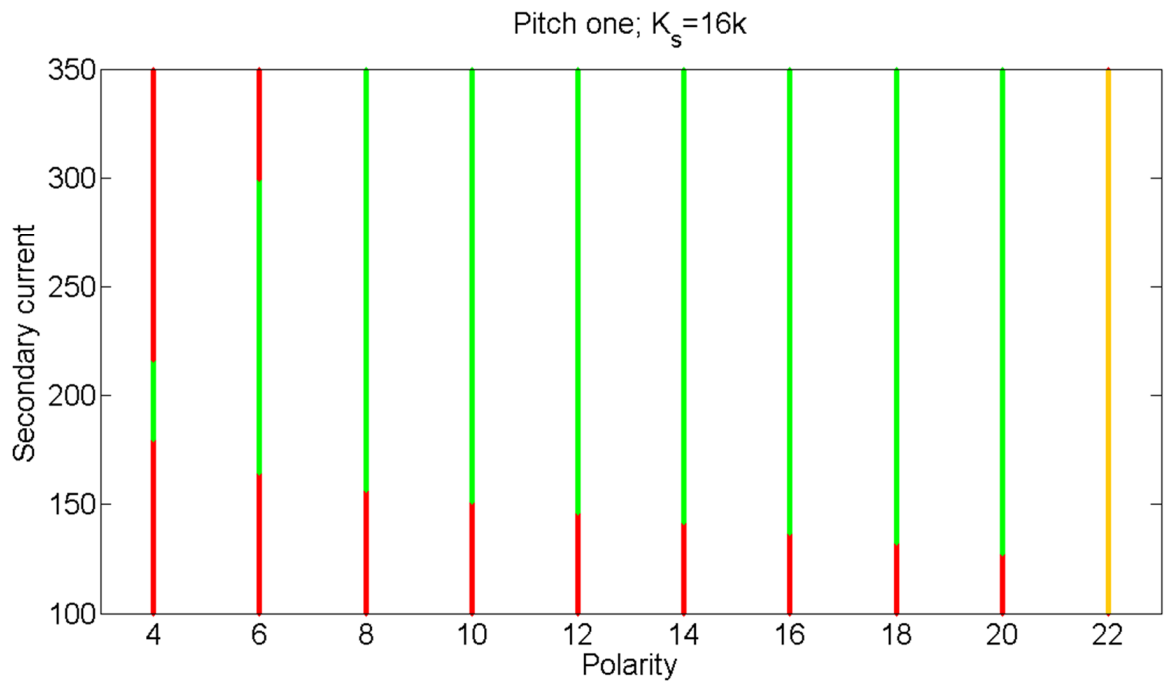

Figure 9 : *Stability diagram with different parameter values.*

For the last Figure, the pitch is raised to 50º. Although the quadrupole becomes unworkable, the sextupole is very much feasible. If the quadrupole bearing is desired then $K_s$ can be made still stronger – a value of 20k A/m is seen to generate a large stable zone at $2n$=4.

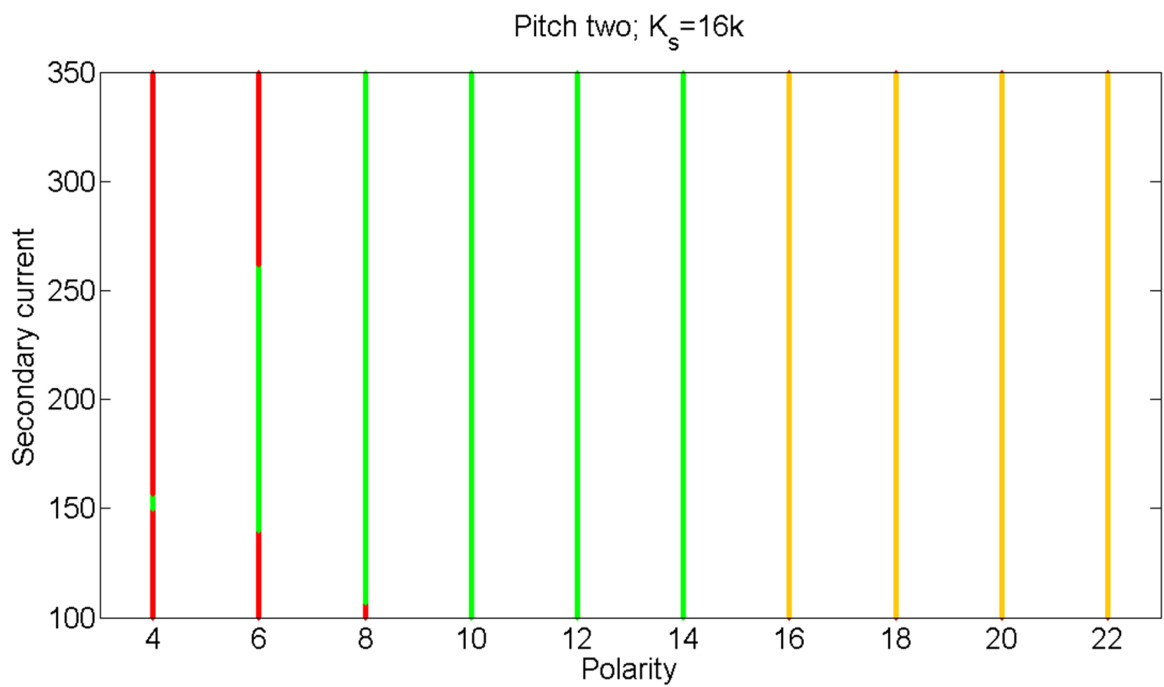

Figure 10 : *Stability diagram with different parameter values.*

At this pitch level, the lift force and torque for the sextupole are 0.72 kgf and 0.31 Nm while those for the octupole are 0.42 kgf and 0.18 Nm respectively. With this figure, the presentation of the stable and unstable regions of operation can be completed.

To improve the stability from oscillatory to decaying type, the natural friction present in the system may be relied on, or additional damping introduced through eddy current mechanisms (vide licet, a replacement of some of the rotor magnetic pole pieces by conducting pieces), viscoelastic elements and aerodynamic spoilers. The calculation of the damping dynamics for a given architecture is a separate problem in its own right and is being reserved for future study.

<div align="center">* * * * *</div>

## Data Accessibility

This paper has no experimental or simulational data.

## Competing Interests

This paper does not create a conflict of interest with any person or agency.

## Author Contributions

I, B SHAYAK, am responsible for the complete process of creation of the entirety of this Article, including but not limited to conception of the idea, physical and mathematical demonstration of effectivity, typing of the manuscript and preparation of the Figures. An error in the original calculation however was spotted and corrected by a Reviewer, mentioned below.

## Acknowledgement

I am grateful to one of the anonymous Reviewers for pointing out an error in the stability analysis in the original version of the manuscript, and for suggesting the GALLINA approach to the problem. I would also like to thank the said as well as the other Reviewers for feedbacks which have resulted in substantial improvement to the quality of presentation of this Article.

# Funding Statement

This research was not funded in whole or in part by any government or private agency.